\documentstyle[psfig,twocolumn,prl,aps,amssymb]{revtex}
\begin{document}
\wideabs{
\draft

\date{\today} \title{Activation energies for spin reversed excitations in the fractional 
quantum Hall effect}
\author{Sudhansu S. Mandal and J.K. Jain}
\address{Department of Physics, 104 Davey Laboratory, The Pennsylvania State
University, University Park, Pennsylvania 16802}
\maketitle

\begin{abstract}

The activation energy measured in transport experiments in the fractional Hall regime  
corresponds to the energy required to create a far separated particle hole pair.
We calculate, for several fractional quantum Hall states, the energy gap for  
the excitation in which the excited composite fermion reverses its spin quantum number.
We consider an ideal system with zero thickness as well as the more realistic quantum well
geometry, while neglecting Landau level mixing and disorder, and find that 
the spin reversed excitations may be relevant for experimentally accessible parameters.

\end{abstract}

}

The phenomenon of the fractional quantum Hall effect (FQHE) \cite {Tsui} occurs  
because of the opening up of gaps at certain fractional filling factors, measured 
experimentally by studying  the temperature dependence of the longitudinal resistance 
\cite{Chang,Willett,Du,Manoharan}).
While the gaps at the integral fillings appear due to the well known quantization of 
the kinetic energy into Landau levels, the FQHE gaps owe their origin to inter-electron 
interaction and signify the appearance of a collective state.  This remarkable correlated state 
with extraordinary properties is described in terms of particles called composite fermions, 
namely electrons bound to an even number of vortices \cite {Jain}.  
The composite fermions experience a reduced effective magnetic field, and 
have an effective filling factor $\nu^*$, which is related to the electron filling factor $\nu$
by $\nu=\nu^*/(2p\nu^*\pm 1)$.  The formation of 
composite fermions gives a simple intuitive explanation for the existence of the gaps.  
A gap opens up when the composite fermions occupy an integral 
number of their Landau levels (called composite-fermion Landau levels), i.e., when  
$\nu^*=n$, which gives FQHE at $\nu=n/(2pn\pm 1)$, 
precisely the observed sequences of fractions.

In addition to the above intuitive understanding, the composite fermion theory gives an 
accurate microscopic description.  Wave functions for the ground state 
and the excitations \cite{Jain} are known, from comparisons with exact results 
obtained in numerical diagonalization studies on finite systems, to be practically identical 
to the exact wave functions \cite{JK}.  In particular, they produce energy gaps with an accuracy of 
a few percent.  The energy gaps have been calculated for several FQHE states
for a strictly two-dimensional electron system with no 
disorder and no Landau level mixing (referred to as the ``ideal" system below) \cite{HR,Fano,JK}, and 
corrections due to finite thickness and Landau 
level mixing have also been estimated \cite{ZDS,Yoshioka,Park,Murthy}. 
While the theory obtains the qualitative trends seen experimentally \cite{Du,Manoharan}, there still 
are significant quantitative disagreements between the theoretical and experimental gaps \cite{Park}. 
Because the composite fermion (CF) theory gives an accurate account of the computer experiments on 
the ideal system, it is believed that the discrepancy between theory and experiment is 
caused by the approximate theoretical treatment of the finite thickness and Landau level 
mixing effects, and also because of the ever present disorder.

This article reports results on the energy gaps to spin reversed excitations.
The gap to creating a spin reversed excitation was studied  
for $\nu=1/3$ prior to the composite fermion theory in exact diagonalization studies 
\cite{Chakraborty,Rezayi}, and it was estimated that for the 
ideal system,  the energy of the spin reversed excitation is 
$\Delta^{\uparrow\downarrow}=\Delta_C^{\uparrow\downarrow} +\Delta_Z$,
with  $\Delta_C^{\uparrow\downarrow}=0.075 V_C $ 
as opposed to the energy of the excitation that does not involve spin reversal
$\Delta^{\uparrow\uparrow}=0.105 V_C$.  Here $\Delta_Z$ is the Zeeman splitting, and the
interaction component is measured in units of $V_C={e^2\over \epsilon l}$ 
where $l=\sqrt{\hbar c/eB}$ is 
the magnetic length and $\epsilon$ is the background dielectric constant.
For GaAs, with the magnetic field quoted in Tesla and the energies in Kelvin,
we have $V_C=50\sqrt{B[T]}$ K and $\Delta_Z=0.30 B[T]$ K, and the above result 
implies that for magnetic fields below $B_c\approx 28$T, the observed 
gap at $\nu=1/3$ corresponds to the spin reversed excitation. 
However, early experimental studies \cite{Furneaux} indicated otherwise.
As we shall see, the crossover magnetic field is a very sensitive function of 
various parameter; the modification of the interaction due to the finite width 
of the actual experimental system 
significantly diminishes the value of the crossover magnetic field and 
the fully polarized excitation becomes relevant for typical parameters.

At first sight, it may seem 
surprising that the interaction energy of the spin reversed excitation is less
than that of fully polarized excitation, because a reasoning based on exchange energy 
considerations would point toward quite the opposite conclusion.  The explanation is that more 
important than exchange energy is the effective Landau level (LL) energy of
composite fermions:  $\Delta^{\uparrow\uparrow}$ involves transition of a composite fermion 
into a {\em higher} CF LL, whereas $\Delta^{\uparrow\downarrow}$ does not. 
We consider here  spin reversed excitations at general filling factors of the form $\nu=n/(2n+1)$.
The difference from $n=1$ case ($\nu=1/3$) is the possibility of transitions into {\em lower} 
CF Landau levels, which might make spin reversed excitations more favorable.  
Our results indicate that spin reversed excitations may be more pervasive than thought earlier;
the gaps at 3/7 and 4/9 may actually correspond to spin reversed excitations for typical 
experiments.

\begin{figure}
\centerline{\psfig{figure=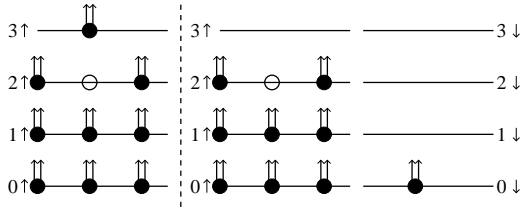,width=3.0in,angle=-90}}
\caption{Schematic depiction of a particle hole pair of composite fermions at $\nu^*=3$ ($\nu=3/7$),
with composite fermions shown as electrons bound to two flux quanta.  The figure to the left of the
vertical dashed line shows the spin conserving excitation, and the figure on the right shows the 
excitation in which a composite fermion falls into a lower spin reversed CF Landau level.  The label 
at the the side of each CF Landau level, e.g., $2\uparrow$ or $1\downarrow$ gives the CF-LL
index along with the spin.  The Zeeman splitting is neglected for simplicity.}
\label{fig1}
\end{figure}

\begin{figure}
\centerline{\psfig{figure=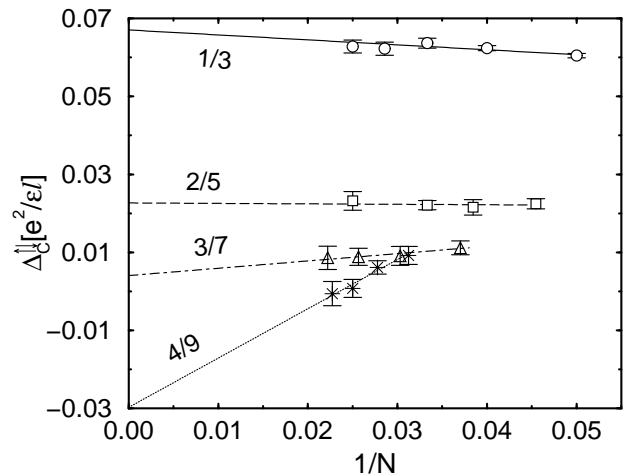,width=4.0in,angle=-90}}
\caption{The determination of the thermodynamic limit by extrapolation of finite system results for 
the quantum well system of width 20nm with electron density $\rho=1.0\times 10^{11}$ cm$^{-2}$.  
A linear fit is assumed. The error bars show the 
statistical uncertainty in the Monte Carlo evaluation of the energy gap. $N$ is the number of
particles, and the energies are given in units of $e^2/\epsilon l$, where $l$ is the magnetic length
and $\epsilon$ is the dielectric constant of the background material.}
\label{fig2}
\end{figure}

We compute the gaps following the method outlined in detail in the literature \cite{JK,Park}, using the 
spherical geometry \cite{Haldane} which considers $N$ electrons on the surface of a sphere, moving
under the influence of a strong radial magnetic field.  The wave function for the ground 
state is given by
$\Psi_{n/(2n+1)}={\cal P}_{LLL}\Phi_n\Phi_1^2$, where $\Phi_n$ is the wave function of the $n$ filled
LL state, and ${\cal P}_{LLL}$ is the lowest LL projection operator. The ground state will be taken to
be fully spin polarized. The wave function for the excited state is 
$\Psi^{\uparrow\downarrow}_{n/(2n+1)}={\cal P}_{LLL}\Phi^{\uparrow\downarrow}_n\Phi_1^2$, 
where $\Phi^{\uparrow\downarrow}_n$ is the wave function of that excited state at $\nu=n$ in which  
one particle has been removed from the highest occupied spin-up Landau level and 
placed in the lowest spin-down Landau level.  The state $\Psi^{\uparrow\downarrow}_{n/(2n+1)}$ is 
then interpreted as the state in which one composite fermions has transferred from highest occupied
spin-up CF-LL to the lowest spin-down CF-LL, as shown in Fig.~(\ref{fig1}).  
We are interested in the energy of this excitation in
limit that the distance between the CF particle and the CF hole is very large,
so we consider the excited state in which they are on the opposite poles of the sphere.
We compute the energy gaps  by evaluating the expectation 
values of the interaction energy ${\hat V}=\sum_{j<k}V(r_{jk})$ in the 
composite fermion wave functions for the
ground and excited states:
$$\Delta_C^{\uparrow\downarrow} =\frac{<\Psi^{\uparrow\downarrow}_{n\over 2n+1}|{\hat V}
|\Psi^{\uparrow\downarrow}_{n\over 2n+1}>}
{<\Psi^{\uparrow\downarrow}_{n\over 2n+1}|\Psi^{\uparrow\downarrow}_{n\over 2n+1}>}
-\frac{<\Psi_{n\over 2n+1}|{\hat V}|\Psi_{n\over 2n+1}>}{<\Psi_{n\over 2n+1}|\Psi_{n\over 2n+1}>}.
$$
The expectation value of the interaction requires evaluation of multidimensional ($2N$-dimensional) 
integrals, which is accomplished by Monte Carlo for different numbers of
particles, and the thermodynamic limit is obtained as shown in Fig.~(\ref{fig2}).
(Prior to an extrapolation of 
our results to the limit of $N\rightarrow \infty$, we
correct for the interaction between the CF particle and the CF hole, which
amounts to a subtraction of $-(2n+1)^{-2}/2\epsilon R$, the 
interaction energy for  
two point-like particles of charges $e/(2n+1)$ and $-e/(2n+1)$ at a distance
$2R$, where $R$ is the radius of the sphere.   We also correct for
a finite size deviation of the 
density from its thermodynamic value, by multiplying by a factor
$\sqrt{\rho/\rho_N}$, where $\rho$ is the thermodynamic density
and $\rho_N$ is the density of the $N$ particle system.)
For the ideal system with zero thickness, we use $V(r)=e^2/\epsilon r$ to obtain the gaps.  
For the more realistic situation, we calculate the profile of the transverse wave function within 
a local density approximation\cite{Stern}, and integrate over the transverse coordinate 
produces an effective 
two-dimensional interaction.  The details of the calculation as well as of 
various approximations made have been outlined in Ref.~\onlinecite{Park}. 

The thermodynamic values for the interaction component of the 
gaps for the ideal system are given in Table I for several filling
factors.  A comparison with $\Delta^{\uparrow\uparrow}$, calculated earlier \cite{Park} and 
reproduced in Table I, produces rather 
large values for the crossover magnetic field, $B_c$, below which the lowest energy gap 
involves spin reversal.  At filling factor $\nu$, the magnetic field is related to the density as  
$\rho=(\nu B[T]/3.9)\times 10^{11}$cm$^{-2}$, from which the densities can be obtained below which 
the gap corresponds to excitation of a spin reversed composite fermion.

Whether the activation gap corresponds to $\Delta^{\uparrow\downarrow}$ or $\Delta^{\uparrow\uparrow}$
depends on whether the ratio $\Delta^{\uparrow\downarrow}/\Delta^{\uparrow\uparrow}$ is smaller 
or larger than one.
As the CF filling $n$ increases, there are two competing effects.  First, the magnitude of the
interaction components of the gaps are expected to decrease, which 
would increase the ratio $\Delta^{\uparrow\downarrow}/\Delta^{\uparrow\uparrow}$ and  
suppress spin reversal.  On the other hand, in creating a spin reversed excitation,
the composite fermion can jump down 
by more CF-Landau levels for larger $n$, thereby leading to a larger decrease in 
the interaction energy relative to the fully polarized excitation, which 
would {\em decrease} the ratio. 
Our microscopic calculation shows that the latter effect wins, at least for the 
ideal situation.

\begin{minipage}{80mm}
\begin{table}[t]
\begin{center}
\begin{tabular}{|c|c|c|c|c|}
$\nu$  & $\Delta_C^{\uparrow\downarrow}=\Delta^{\uparrow\downarrow}-\Delta_Z$ 
[${e^2\over \epsilon l}$]  
&   $\Delta^{\uparrow\uparrow}$ [${e^2\over \epsilon l}$]  & $B_c$[T] & $B_{tr}$[T] \\ \hline
1/3 & 0.0740(24) &  0.106(3) &        28 & - \\ \hline
2/5 & 0.0235(55) &  0.058(5)  &       33  & 3.5 \\ \hline
3/7 & 0.0022(94) &  0.047(4)  &       56  & 5.8 \\ \hline
4/9 & $- 0.0402(135)$ &  0.035(6)  &     157 & 7.4 \\
\end{tabular}
\end{center}
\caption{The interaction components of the
lowest excitation gaps involving spin reversal
($\Delta_C^{\uparrow\downarrow}=\Delta^{\uparrow\downarrow}-\Delta_Z$) and no spin reversal
$\Delta^{\uparrow\uparrow}$. The magnetic field is quoted in Tesla and the
energy is quoted in units of $e^2/\epsilon l$, where $l$ is the
magnetic length and $\epsilon$ is the dielectric constant of the background material.
The results are for $V(r)=e^2/\epsilon r$, as appropriate for a system with zero thickness;
Landau level mixing and disorder are not considered.
The crossover magnetic field $B_c$ is given for parameters appropriate
for GaAs; the spin reversed excitation has the lowest energy for fields below $B_c$.
Also given is the magnetic fields $B_{tr}$, at which
a transition into a partially polarized ground state takes place.
The gaps $\Delta_C^{\uparrow\uparrow}$ are taken from Jain and Kamilla
\protect \cite{JK}, $B_{tr}$ from Park and Jain \protect  \cite{Park2}, and the value of
$\Delta_C^{\uparrow\downarrow}$ for $\nu=1/3$ is consistent with that of Rezayi
\protect\cite{Rezayi}.
\label{tab:Tab1}}
\end{table}
\end{minipage}

Our calculation above assumes a fully polarized ground state, but 
it is well known, from numerical diagonalization studies \cite{Zhang}, from the composite fermion theory
\cite{Wu,Park2}, and also experimentally \cite{Clark,Du2},
that for sufficiently small Zeeman energies, the FQHE ground state is not fully
polarized.  The fully polarized state at $\nu^*=n$, denoted by $(n,0)$, makes a transition 
into the state 
$(n-1,1)$, which has $n-1$ spin-up and 1 spin-down CF-LLs occupied, as the Zeeman energy is reduced.  
Because a fraction $1/n$ of the total number of particles reverses its spin at the transition, the 
magnetic field $B_{tr}$ at 
the transition is determined from the equation $E_C^{(n,0)}-E_C^{(n-1,1)}=\Delta_Z/n$,
where $E_C$ is the interaction energy per particle.
For the ideal case,  the energy differences\cite{Park2}
 $E_C^{(n,0)}-E_C^{(n-1,1)}$ for $\nu^*=2$, 3, and 4 are  
0.0056, 0.0048, 0.0041 in units of $V_C$, with the resulting $B_{tr}$ given in Table I. 
There is a range of magnetic fields with $B_{tr}< B < B_c$ where the ground state is fully
polarized but the lowest energy excitation involves spin reversal.

\begin{figure}
\centerline{\psfig{figure=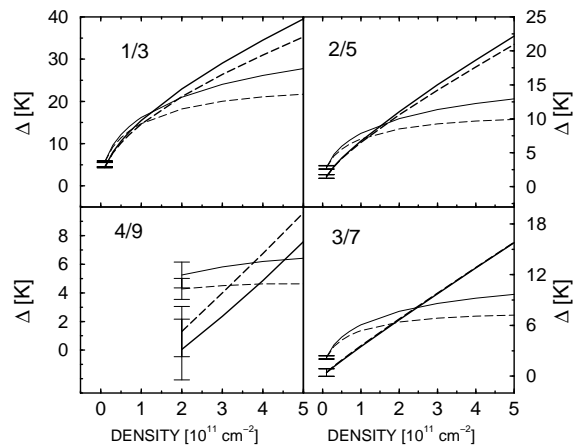,width=3.5in,angle=-90}}
\caption{The energies of spin reversed (thick lines) and spin polarized (thin lines) 
excitations, in Kelvin, for several filling factors, as a function of the electron 
density, for quantum
wells of widths 20 nm (solid lines) and 30 nm (dashed lines).  The gaps include the 
Zeeman contribution, wherever appropriate; the gap $\Delta^{\uparrow\downarrow}$ at $\nu=3/7$
is essentially equal to the Zeeman energy
because of the smallness of the Coulomb contribution, $\Delta_C^{\uparrow\downarrow}$.
The typical Monte Carlo uncertainty is shown at the left. 
The energies of the spin polarized excitations 
are taken from Park, Meskini, and Jain \protect \cite{Park}.}
\label{fig3}
\end{figure}

These numbers might suggest that the gap to spin reversed excitation
is the relevant gap for GaAs for almost 
all of the experimental range of parameters.  That is not the case, however, because  
$B_c$ is a sensitive function of certain features left out in the ideal model.  We consider now the 
finite thickness corrections, that will reduce the interaction components of 
both $\Delta^{\uparrow\downarrow}$ and $\Delta^{\uparrow\uparrow}$, but leave the Zeeman contribution
to  $\Delta^{\uparrow\downarrow}$ unaffected.
The calculated gaps are given in Fig.~(\ref{fig3}) for quantum well widths 
20 and 30 nm as a function of the density.  The crossover magnetic fields are lowered 
compared to the ideal case, but are still in experimentally accessible regime for all 
of the filling factors considered.

As is clear in the above, the crossover magnetic field is a sensitive function of the {\em difference}
between the gaps, and a reliable estimation of its value would require an accurate 
quantitative understanding of
the observed gaps, which is not available at the present.  The treatment of finite thickness is 
reliable only on the order of 20\% for each individual gap.
Also, we have not considered in this work the effect of LL mixing \cite{LLmixing}, 
which further modifies the gap, perhaps by 10-20\% for typical experimental parameters.  
Even after both the finite thickness and LL mixing corrections are included in the theory, the 
actual values of the theoretical and experimental gaps are off by approximately a factor of
two \cite{Park}, presumably because of disorder.   
Due to the unsatisfactory quantitative understanding of the observed gaps, it is not possible
to say how trustworthy our estimates of $B_{c}$ are insofar as the actual experiments are 
concerned, but the qualitative trends found in our study ought to be robust.

This work was supported in part by the
National Science Foundation under grant no. DMR-9986806.

\end{document}